\documentclass[conference]{IEEEtran}
\usepackage{amsfonts}
\usepackage{stmaryrd}
\usepackage{bbding}
\usepackage{algorithm}
\usepackage{algorithmic}
\makeatletter
\def\ps@headings{%
\def\@oddhead{\mbox{}\scriptsize\rightmark \hfil \thepage}%
\def\@evenhead{\scriptsize\thepage \hfil \leftmark\mbox{}}%
\def\@oddfoot{}%
\def\@evenfoot{}}
\makeatother 

\usepackage{amsthm}

\theoremstyle{plain}
\newtheorem{mydef}{Definition}
\newtheorem{mytheorem}{Theorem}
\newtheorem{mylemma}{Lemma}

\ifCLASSINFOpdf
\else
  \usepackage[dvips]{graphicx}
\fi
\usepackage[cmex10]{amsmath}
\begin{document}
\title{PS-TRUST: Provably Secure Solution for Truthful Double Spectrum Auctions}


\author{\IEEEauthorblockN{Zhili Chen, Liusheng Huang, Lu Li, Wei Yang, Haibo Miao, Miaomiao Tian, Fei Wang}
\IEEEauthorblockA{School of Computer Science and Technology, University of Science and Technology of China, Hefei, China\\
Suzhou Institute for Advanced Study, University of
Science and Technology of China, Suzhou, China\\
\small{Email: \{zlchen3, lshuang\}@ustc.edu.cn,
liluzq@mail.ustc.edu.cn, qubit@ustc.edu.cn, \{mhb, miaotian,
scuwf\}@mail.ustc.edu.cn}}}

\maketitle

\begin{abstract}

Truthful spectrum auctions have been extensively studied in recent
years. Truthfulness makes bidders bid their true valuations,
simplifying greatly the analysis of auctions. However, revealing
one's true valuation causes severe privacy disclosure to the
auctioneer and other bidders. To make things worse, previous work on
secure spectrum auctions does not provide adequate security. In this
paper, based on TRUST, we propose PS-TRUST, a provably secure
solution for truthful double spectrum auctions. Besides maintaining
the properties of truthfulness and special spectrum reuse of TRUST,
PS-TRUST achieves provable security against semi-honest adversaries
in the sense of cryptography. Specifically, PS-TRUST reveals nothing
about the bids to anyone in the auction, except the auction result.
To the best of our knowledge, PS-TRUST is the first provably secure
solution for spectrum auctions. Furthermore, experimental results
show that the computation and communication overhead of PS-TRUST is
modest, and its practical applications are feasible.
\end{abstract}


% \fi
\IEEEpeerreviewmaketitle

\section{Introduction}

As the rapid development of wireless technologies, the scarcity of
radio spectrum attracts more and more attention. Under the
traditional static spectrum allocation scheme by government, the
utilization of the radio spectrum is severely inefficient. Many
spectrum channels are idle most of the time under their current
owners, whereas ever-increasing new wireless users are starving for
spectrum. Therefore, spectrum redistribution is highly significant
for improving the overall spectrum utilization and thus alleviating
the problem of spectrum scarcity. Open markets for spectrum
redistribution, such as Spectrum Bridge \cite{bridges}, have already
appeared to provide services for buying, selling, and leasing idle
spectrum.

As a well-known approach to spectrum redistribution, spectrum
auctions are preferred by people for its fairness and allocation
efficiency. In recent years, there have been extensive studies on
spectrum auctions, most of which achieve truthfulness to make
bidders reveal their true valuations of spectrum channels. However,
revealing one's true valuation causes severe privacy disclosure.
Literature \cite{11panm} illustrated two vulnerabilities of truthful
auctions, i.e. frauds of the insincere auctioneer, and bid-rigging
between the auctioneer and the bidders, in which the auctioneer
takes advantage of the knowledge of bidders' bids. Furthermore, when
one bidder knows other bidders' bids after an auction, he will
probably not bid his true valuation in repeated auctions. That is,
an original truthful auction will probably become untruthful when
repeated, due to the revelation of all bidders' bids in the previous
auctions \cite{07mcsherryf}. Therefore, protecting the privacy of
bidders is of great importance.

There have been many researches on privacy preserving auctions, such
as \cite{99naorm}\cite{02pengk}\cite{03suzukik}\cite{04yokoom}.
However, spectrum is quite different from traditional goods, for it
can be well reused in both spatial and time dimensions. Thus,
traditional privacy preserving auctions cannot be directly applied
to spectrum auctions. Recently, some works about privacy preserving
spectrum auctions have also been proposed
\cite{11panm}\cite{13huangq}. These works dealt with only
single-sided spectrum auctions. Furthermore, they fell short of
providing adequate security. In the sense of cryptography, a
protocol is secure implies that no participating party can learn any
information beyond the output of the protocol. However, both the two
approaches reveal some information that cannot be inferred from the
outputs. For example, in \cite{11panm}, the auctioneer can easily
get the sums of bids for all the possible allocations for each
subnetwork by decrypting $\textbf{E}_{\xi}$; in \cite{13huangq}, the
auctioneer gets to know the bids of all buyer groups and their
ranking order in the auctions. The information mentioned above is
more than the auction result, which normally includes the winner set
and the pricing information.

\begin{figure}[!t]
\centering
\includegraphics[width=0.4\textwidth]{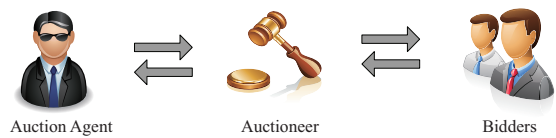}
\caption{Auction Framework for PS-TRUST} \label{fig:framework}
\end{figure}

In this paper, we propose PS-TRUST, a provably secure solution for
truthful double spectrum auctions. The auction framework of PS-TRUST
is shown as in Fig. \ref{fig:framework}. This framework introduces
an auction agent who cooperates with the auctioneer to securely
compute the auctions. Neither the auctioneer nor the auction agent
is a trusted party, but they are assumed not to collude with each
other. Furthermore, we restrict that bidders can only communicate
with the auctioneer, keeping the communication pattern simple and
identical to that of an insecure auction. PS-TRUST reveals nothing
but the auction result including the selling and buying clearing
prices, and the seller and buyer winner sets. The main contributions
can be summarized as follows.

(1) We design PS-TRUST based on homomorphic encryption schemes. By
representing the bids in encrypted bit vectors (EBVs), we design
secure algorithms for addition, constant multiplication, and
maximum/minimum selection for EBV bids. And then, based on these
algorithms, we present a secure auction procedure, which reveals
nothing about the bids except the auction result.

(2) We apply the definition of security against semi-honest
adversaries to formally prove the security of PS-TRUST. To the best
of our knowledge, this is the first work to formally prove the
security, in the sense of cryptography, of a solution to spectrum
auctions.

(3) We analyze the computation and communication complexities of
PS-TRUST, implement it in Java to evaluate running times and message
volumes, and conclude that its computation and communication
overhead is modest.

The remainder of this paper is organized as follows. In Section
\ref{sec:relatedwork}, a brief review of related work is given. In
Section \ref{sec:problemstatement}, we describe the problem
statement. Next, we provide some preliminaries in Section
\ref{sec:preliminaries}. In Section \ref{sec:PS-TRUST}, we present
the detailed design of PS-TRUST, and prove formally its security.
Then, in Section \ref{sec:experiment}, we implement PS-TRUST,
analyze and evaluate its computation and communication overhead.
Finally, we conclude our work in Section \ref{sec:conclusion}.

\section{Related Work}\label{sec:relatedwork}

Spectrum auctions have been studied extensively in recent years. For
instance, Zhou et al. proposed VERITAS \cite{08zhoux}, a
single-sided truthful spectrum auction supporting diverse bidding
formats. Zhou et al. proposed TRUST \cite{09zhoux}, the first
truthful double spectrum auction framework enabling spectrum reuse.
Deek et al. proposed Topaz \cite{11deeklb} to tackle time-based
cheating in online spectrum auctions. Al-Ayyoub and Gupta
\cite{11al-ayyoubm} designed a polynomial-time truthful spectrum
auction mechanism with a performance guarantee on revenue. Xu et al.
\cite{11xup-a}\cite{11xup-b} proposed efficient online spectrum
allocations in multi-channel wireless networks. TAHES \cite{12fengx}
addressed the issue of heterogeneous spectrums in truthful double
spectrum auctions. Dong et al. \cite{12dongm} tackled the spectrum
allocation problem with time-frequency flexibility in cognitive
radio networks via combinatorial auction. However, most of the
existing spectrum auction mechanisms do not provide any guarantee of
security.

Extensive work has focused on privacy preserving auction design in
the past decade. Brandt and Sandholm \cite{08brandtf} investigated
unconditional full privacy in sealed-bid auctions. In \cite{99naorm}
\cite{02pengk}\cite{03suzukik}\cite{04yokoom} the authors employed
various cryptography techniques to achieve security in diverse
auction schemes. Unfortunately, when applied to spectrum auctions,
these traditional privacy preserving auctions either require
exponential complexity, or lead to significant degradation of
spectrum utilization. Recently, papers \cite{11panm} and
\cite{13huangq} provide solutions for privacy preserving spectrum
auctions, but they only addressed single-sided spectrum auctions.
What is more, as mentioned above, they fell short of providing
security in the sense of cryptography.

\section{Problem Statement}\label{sec:problemstatement}

\subsection{Auction Problem}

We consider a double spectrum auction, which is single-rounded with
one auctioneer $\mathcal{A}$, a seller set $\mathbb{S} = \{s_1,
s_2,$ $...,s_M\}$, and a buyer set $\mathbb{B} = \{b_1, b_2,...,
b_N\}$. In the auction, each seller $s_i$ contributes exactly one
channel and each buyer $b_j$ requests only one channel. The channels
are homogenous to buyers so that their requests are not channel
specific. Each channel contributed by sellers can potentially be
reused by multiple non-conflicting buyers who are separated far
enough.

\subsection{TRUST}

TRUST \cite{09zhoux} has provided a truthful framework for this
double spectrum auction problem, with spatial spectrum reuse being
well exploited. Since TRUST \cite{09zhoux} is based on McAfee's
double auction design, we briefly review both of them.

\subsubsection{McAfee's Design}

McAfee's design of double auctions is most widely used
\cite{92mcafeerp}, which achieves economic properties of
truthfulness, individual rationality, and ex-post budget balance.
This design assumes that there are $M$ sellers and $N$ buyers, and
all goods auctioned are homogenous. Each seller $s_i$ bids $v^s_i$
to sell a good, and each buyer $b_j$ bids $v^b_j$ to buy a good. The
auction proceeds as follows:

(1) Bid sorting: Sort bids of sellers in non-decreasing order and
bids of buyers in non-increasing order:
\[
\begin{array}{l}
v^s_1 \le v^s_2 \le ... \le v^s_M \\
v^b_1 \le v^b_2 \le ... \le v^b_N
\end{array}
\]

(2) Winner determination: Find $k = arg \max{\{v^s_k \le v^b_k\}}$,
the index of the last profitable transaction. Then the first $(k-1)$
sellers and the first $(k-1)$ buyers are the auction winners.

(3) Pricing: Pay each winning seller equally by $v^s_k$, and charge
each winning buyer equally by $v^b_k$.

\subsubsection{TRUST Design}

TRUST followed the methodology of McAfee's design, and enabled
spectrum spatial reuse. It consists of the following three steps:

(1) Buyer group formation: form non-conflicting buyer groups based
on buyers' conflict graph but independent of their bids.

(2) Winner determination: Each buyer group bids a value obtained by
multiplying its smallest buyer bid with its size, and acts as a
single ``buyer''. Then the auctioneer applies just the same winner
determination as that of the McAfee's design, resulting in that the
first $(k-1)$ sellers and the buyers in the first $(k-1)$ buyer
groups are the auction winners.

(3) Pricing: Pay each winning seller equally by the $k^{\text{th}}$
seller bid, and charge each buyer group equally by the
$k^{\text{th}}$ buyer group bid, which is evenly shared among the
buyers in the group.

\subsection{Securing TRUST}

As described above, TRUST has provided a good solution to the
auction problem mentioned. However, in TRUST, no security issues are
considered, and all bids are completely exposed to the auctioneer,
and even to all bidders. This could result in the following two
problems: (1) A dishonest auctioneer could temper the auction result
to increase his utility \cite{11panm}; (2) The knowledge of the
historical true valuations of other bidders could make one bidder
conceal his true valuation in a repetition of a truthful auction
\cite{07mcsherryf}.

In this work, our aim is to secure TRUST by protecting the privacy
of bidders, i.e., their bids. However, how to correctly compute the
auction while reveal nothing about the bids beyond the auction
result (including selling and buying clearing prices, seller and
buyer winner sets) in the auction process, is challenging.
Furthermore, how to prove the security in the sense of cryptography
is non-trivial, too.

\section{Preliminaries}\label{sec:preliminaries}

In this section, we introduce some preliminaries for the design of
PS-TRUST.

\subsection{Security Formulation}

\begin{figure}[ht]
\centering
\includegraphics[width=0.5\textwidth]{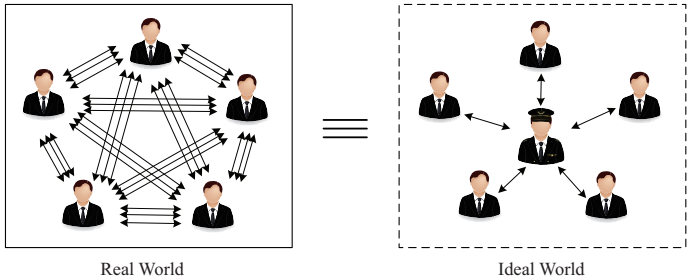}
\caption{The Security Formulation of Ideal/Real Simulation Paradigm}
\label{fig:serurityform}
\end{figure}

In cryptography area, the standard security formulation is called
ideal/real simulation paradigm \cite{04goldreicho} \cite{10Hazayc},
as shown in Fig. \ref{fig:serurityform}. In this formulation, a real
protocol execution in the ``real world'' is mapped to an ideal
functionality calling in the ``ideal world''. In the ideal world,
there is an external trusted (and incorruptible) party willing to
help the parties carry out their computation. The ideal
functionality calling means that the parties simply send their
inputs to the trusted party, which computes the desired
functionality and passes each party its prescribed output. While, in
the real world, there is no external trusted party, and the real
protocol execution means the parties run the protocol amongst
themselves without any help. We say that a protocol is \emph{secure}
if its real protocol execution emulates its ideal functionality
calling. That is, no adversary can do more harm in its real protocol
execution than in its ideal functionality calling. However,
successful adversarial attacks cannot be performed in the ideal
functionality calling. We therefore conclude that all adversarial
attacks on the real protocol execution must also fail for a secure
protocol.

\subsection{Adversarial Models}

Under the security formulation of ``ideal/real simulation
paradigm'', the adversarial models can be classified as semi-honest
adversarial model and malicious adversarial model
\cite{04goldreicho}.

In semi-honest adversarial model, even a corrupted party correctly
follows the protocol specification. However, the adversary obtains
the internal state of all the corrupted parties, and attempts to use
this to learn information more than the output. This adversarial
model may be used in settings where running the ``correct'' protocol
can be enforced. Semi-honest adversaries are also called
``honest-but-curious adversaries'' and ``passive adversaries''.

In malicious adversarial model, the corrupted parties can
arbitrarily deviate from the protocol specification, according to
the adversary's instructions. Security against malicious adversaries
is so strong that it ensures that no adversarial attack can succeed.
Malicious adversaries are also called ``active adversaries''.

Although protocols secure against malicious adversaries exist
theoretically, they are far too inefficient to implement. So, in
this paper, we apply semi-honest adversarial model for the cause of
practical applications. Specifically, in our context, we assume that
the auctioneer and the auction agent follow the auction protocol
specification, but one of them could act as a semi-honest adversary.
The adversary obtains the internal state of the auction, and
attempts to learn information about the bids beyond the auction
result.

\subsection{Paillier Cryptosystem}

In order to achieve the security of spectrum auctions, a
semantically secure cryptosystem is needed. In our design,
Paillier's homomorphic cryptosystem $(G, E, D)$ is applied, where
$G$, $E$ and $D$ denote the key generation algorithm, encryption
algorithm, and decryption algorithm, respectively. The properties of
a Paillier cryptosystem include homomorphic addition,
indistinguishability, and self-blinding \cite{99paillierp}:

\textbf{(1) Homomorphic addition}: The product of two ciphertexts
will decrypt to the sum of their corresponding plaintexts, and the
$k^{\text{th}}$ power of a ciphertext will decrypt to the product of
$k$ and its corresponding plaintext.
\begin{small}
\[
D(E(m_1, r_1) \cdot E(m_2, r_2) \text{ mod } N^2) = m_1 + m_2 \text{
mod } N
\]
\[
D(E(m, r)^k \text{ mod } N^2) = k \cdot m \text{ mod } N
\]
\end{small}
where $N$ is the product of two large primes, which is public to
users, and $r_1$, $r_2$ and $r$ are random numbers.

\textbf{(2) Indistinguishability:} If the same plaintext $m$ is
encrypted twice, these two ciphertexts $E(m, r_1)$ and $E(m, r_2)$
are totally different, and no one can succeed in distinguishing them
with a significantly higher probability than random guess without
decrypting them.

\textbf{(3) Self-blinding}: Any ciphertext can be publicly changed
into another one without affecting the plaintext. This means that a
randomized chipertext $E(m, r^\prime)$ can be computed from the
ciphertext $E(m, r)$ without knowing eight the decryption key or the
original plaintext.

\section{PS-TRUST}\label{sec:PS-TRUST}

In this section, we present the design of PS-TRUST. We first
describe the secure bid representation and operations, then present
the detailed secure auction design, and finally prove formally that
PS-TRUST is secure against semi-honest adversaries.

\subsection{Secure Bid Representation and Operations}

In PS-TRUST, we use encrypted bit vectors to securely represent
bids.

\begin{mydef}[Encrypted Bit Vector] \label{def:ebv}
The Encrypted Bit Vector (EBV) representation of value $v$ is a
vector $\textbf{e}(v)$ of ciphertexts like
\begin{small}
\begin{equation}\label{equ:bvr}
\textbf{e}(v) = (e_1, e_2, ..., e_K) = (E(\sigma_1),E(\sigma_2),
...,E(\sigma_K))
\end{equation}
\end{small}
where $E(.)$ is Paillier's encryption function, $K$ is the bit
length, $(\sigma_1, \sigma_2,..., \sigma_K)$ denotes the binary
representation of $v$, with $\sigma_1$ the most significant bit, and
$\sigma_K$ the least significant bit.
\end{mydef}

With the definition of EBV, we can develop secure algorithms for EBV
bid operations including addition, constant multiplication, and
minimum/maximum selection. With these algorithms, the algorithm
runner (AR) without the secret key can compute the corresponding bid
operations on EBV bids, and get an encrypted result, knowing nothing
about the bids. Then this encrypted result can be used as either an
intermediate result for further computations or a part of the final
output decrypted by the key holder (KH) with the secret key.

But how do we compute on EBV bids? Due to the homomorphic addition,
addition of two values in $\mathbb{Z}_N$ can be computed directly by
multiplying their ciphertexts, while multiplication can be computed
with the help of the KH who can do decryption using Protocol
\ref{protocol:product}. Furthermore, the XOR operation signified by
$\oplus$ can be turned into additions and multiplications in
$\mathbb{Z}_N$ by the fact that:
\begin{equation}\label{equ:xor}
c \oplus d = c + d - 2cd
\end{equation}
Thus, to design the secure algorithms for the operations on EBV
bids, we only need to turn all operations into additions and
multiplications in $\mathbb{Z}_N$, and XOR operations.

\floatname{algorithm}{Protocol}

\begin{algorithm}[htb]
\caption{Product of Two Numbers in $\mathbb{Z}_n$}
\label{protocol:product}
\begin{algorithmic}[1]

\REQUIRE ~~\\
    AR holds $E(x)$ and $E(y)$\\

\ENSURE ~~\\
    AR holds $E(xy)$

\textbf{Step AR1:}

\STATE $x_1 \in_R \mathbb{Z}_n$; $y_1 \in_R \mathbb{Z}_n$; // Select
randomly

\STATE $E(x_2) = E(x).E(-x_1)$; // $x_2=x-x_1 \text{ mod } N$;

\STATE $E(y_2) = E(y).E(-y_1)$; // $y_2=y-y_1 \text{ mod } N$;

\STATE Sends $E(x_2)$ and $E(y_2)$ to AA;

\textbf{Step KH2:}

\STATE $x_2 = D(E(x_2))$; $y_2 = D(E(y_2))$;

\STATE Sends $E(x_2y_2)$ to AE;

\textbf{Step AR3:}

\STATE $E(xy)=E(x_1y_1) \cdot E(y_2)^{x_1} \cdot E(x_2)^{y_1} \cdot
E(x_2y_2)$;

\end{algorithmic}
\end{algorithm}

According to the discussion above, the secure algorithms for EBV bid
addition and EBV bid constant multiplication are straightforward,
and are shown in Algorithms \ref{protocol:add} and
\ref{protocol:mul}, respectively.

\renewcommand{\algorithmicrequire}{\textbf{Input:}}  %Use Input in the format of Algorithm
\renewcommand{\algorithmicensure}{\textbf{Output:}}  %UseOutput in the format of Algorithm

%\floatname{algorithm}{Protocol}

\begin{algorithm}[htb]
\caption{EBVAdd$(\textbf{e}(v^A), \textbf{e}(v^B))$}
\label{protocol:add}
\begin{algorithmic}[1]
\REQUIRE ~~\\
    EBV bids $\textbf{e}(v^A)$ and $\textbf{e}(v^B)$\\

\ENSURE ~~\\
    Sum $\textbf{e}(v^{AB})$

\STATE Compute Line \ref{line:addstart} to \ref{line:addend} over
encrypted bits $E(\sigma^A_i)$ and $E(\sigma^B_i)$, where $1 \le i
\le K$, using homomorphic properties and Protocol
\ref{protocol:product}.

// For clarity, we describe these lines by plain bits.

\STATE\label{line:addstart} $\sigma^{AB}_K = \sigma^A_K \oplus
\sigma^B_K$; $c^{AB}_K = \sigma^{A}_K \cdot \sigma^{B}_K$;

\FOR{($i = K - 1$; $i >= 1$; $i = i - 1$)}

\STATE $\sigma^{AB}_i = \sigma^A_i \oplus \sigma^B_i \oplus
c^{AB}_{i+1}$;

\STATE $c^{AB}_i = \sigma^A_i \cdot \sigma^B_i \oplus \sigma^A_i
\cdot c^{AB}_{i+1} \oplus \sigma^B_i \cdot c^{AB}_{i+1}$;

\ENDFOR \label{line:addend}

\STATE $\textbf{e}(v^{AB}) = (E(\sigma^{AB}_1), E(\sigma^{AB}_2),
..., E(\sigma^{AB}_K))$;

\RETURN{$\textbf{e}(v^{AB})$;}

\end{algorithmic}
\end{algorithm}

\begin{algorithm}[htb]
\caption{EBVMul$(\textbf{e}(v), n)$} \label{protocol:mul}
\begin{algorithmic}[1]

\REQUIRE ~~\\
    EBV bid $\textbf{e}(v)$ and integer $n = (\sigma^{(n)}_1, \sigma^{(n)}_2, ..., \sigma^{(n)}_K)$\\

\ENSURE ~~\\
    Product $\textbf{P} = \textbf{e}(n \cdot v)$

\STATE $\textbf{P} = \textbf{e}(0)$;

\FOR{($i = 1$; $i <= K$; ++$i$)}

\IF{($\sigma^{(n)}_i == 1$)}

\STATE $\textbf{P}^* = \textbf{e}(v)$ shifted left $(K - i)$ bits;

\STATE $\textbf{P} =$ EBVAdd$(\textbf{P}, \textbf{P}^*)$;

\ENDIF

\ENDFOR

\RETURN{$\textbf{P}$;}

\end{algorithmic}
\end{algorithm}

Now, we design secure algorithms for minimum selection. We first
consider the two-bid case. Suppose that the AR holds two EBV bids,
denoted by
\begin{equation}\label{equ:abbvr}
\begin{small}
\begin{array}{l}
e(v^A) = (E(\sigma^A_1),E(\sigma^A_2), ...,E(\sigma^A_K)) \text{, and}\\
e(v^B) = (E(\sigma^B_1),E(\sigma^B_2), ...,E(\sigma^B_K))
\end{array}
\end{small}
\end{equation}
It can compute the location of the minimum bid as
\begin{equation}\label{equ:abmin}
\begin{small}
\begin{array}{l}
R^{min}_{AB} = (\sigma^A_1 \oplus \sigma^B_1)\sigma^A_1 +
(\sigma^A_1 \oplus \sigma^B_1 \oplus 1)(\sigma^A_2 \oplus
\sigma^B_2)\sigma^A_2+\\
~~~~~(\sigma^A_1 \oplus \sigma^B_1 \oplus 1)(\sigma^A_2 \oplus
\sigma^B_2 \oplus 1)(\sigma^A_3 \oplus \sigma^B_3)\sigma^A_3+...+\\
~~~~~(\sigma^A_1 \oplus \sigma^B_1 \oplus 1)...(\sigma^A_{K-1}
\oplus \sigma^B_{K-1} \oplus 1)(\sigma^A_K \oplus
\sigma^B_K)\sigma^A_K
\end{array}
\end{small}
\end{equation}
on the encrypted bits, where $R^{min}_{AB}$ is defined as
\begin{equation}\label{equ:minres}
\begin{small}
R^{min}_{AB} = \left\{
\begin{array}{l}
0,~if~v^A \le v^B\\
1,~if~v^A > v^B
\end{array}\right.
\end{small}
\end{equation}

Therefore, we can design the secure algorithm for two-bid minimum
selection as shown in Algorithm \ref{protocol:2mbs}. Note that the
order of the two bids matters in the result. If the two bids are
equal, the first one is picked up.

\renewcommand{\algorithmicrequire}{\textbf{Input:}}  %Use Input in the format of Algorithm
\renewcommand{\algorithmicensure}{\textbf{Output:}}  %UseOutput in the format of Algorithm
\floatname{algorithm}{Algorithm}

\begin{algorithm}[htb]
\caption{TwoBidMin$(\textbf{e}(v^A), \textbf{e}(v^B))$}
\label{protocol:2mbs}
\begin{algorithmic}[1]

\REQUIRE ~~\\
    EBV bids $\textbf{e}(v^A)$ and $\textbf{e}(v^B)$\\

\ENSURE ~~\\
    Comparison result $E(R^{min}_{AB})$ and minimum bid $\textbf{e}(v^{AB})$

\STATE Compute Line \ref{line:start} to \ref{line:end} over
encrypted bits $E(\sigma^A_i)$ and $E(\sigma^B_i)$, where $1 \le i
\le K$.

// For clarity, we describe these lines by plain bits.

\FOR{($i = 1$; $i <= K$; ++$i$)}\label{line:start}

\STATE $x^{AB}_i = \sigma^A_i \oplus \sigma^B_i$; $x^{AB*}_i =
x^{AB}_i \oplus 1$;

\ENDFOR

\STATE $R^{min}_{AB} = x^{AB}_1 \cdot \sigma^A_1$; $R = 1$;

\FOR{($i = 2$; $i <= K$; ++$i$)}

\STATE $R = R \cdot x^{AB*}_{i-1}$;

\STATE $R^{min}_{AB} = R^{min}_{AB} + R \cdot x^{AB}_i \cdot
\sigma^A_i;$

\ENDFOR\label{line:end}

\STATE Compute $\textbf{e}(v^{AB}) = (E(\sigma^{AB}_1),
E(\sigma^{AB}_2),...,E(\sigma^{AB}_K))$, where $\sigma^{AB}_j =
\sigma^A_j \cdot (1-R^{min}_{AB}) + \sigma^B_j \cdot R^{min}_{AB}$,
$1 \le j \le K$;

\RETURN{$(E(R^{min}_{AB}), \textbf{e}(v^{AB}))$;}

\end{algorithmic}
\end{algorithm}

Based on Algorithm \ref{protocol:2mbs}, we can develop the algorithm
for multi-bid minimum selection as shown in Algorithm
\ref{protocol:nmbs}.

\begin{algorithm}[htb]
\caption{MultiBidMin$(E(\mathbb{B}))$} \label{protocol:nmbs}
\begin{algorithmic}[1]

\REQUIRE ~~\\
    EBV bids $E(\mathbb{B}) = \{\textbf{e}(v^i) | 1 \le i \le n\}$\\

\ENSURE ~~\\
    Comparison result $E(R^{min}_{1,n})$ and minimum bid $\textbf{e}(v^{n*})$\\

\STATE $(E(R^{min}_{1,2}), \textbf{e}(v^{2*}))=$
TwoBidMin$(\textbf{e}(v^1), \textbf{e}(v^2))$;

\FOR{($i = 2$; $i < n$; ++$i$)}

\begin{small}
\STATE $(E(R^{min}_{i*,i+1}), \textbf{e}(v^{(i+1)*}))=$
TwoBidMin$(\textbf{e}(v^{i*}), \textbf{e}(v^{i+1}))$;
\end{small}

\STATE $E(R^{min}_{1,i+1}) = E(R^{min}_{1,i} \cdot (1 -
R^{min}_{i*,i+1}) + (i + 1) \cdot R^{min}_{i*,i+1})$;

\ENDFOR

\RETURN{$(E(R^{min}_{1,n}), \textbf{e}(v^{n*}))$;}

\end{algorithmic}
\end{algorithm}

In Algorithm \ref{protocol:nmbs}, the inputs are EBV bids of a set
of bidders indexed from $1$ to $n$. $\textbf{e}(v^{i*})$ represents
the minimum EBV bid of bidders from bidders $1$ to $i$.
$R^{min}_{i*,i+1}$ denotes the comparison result of  the minimum bid
of bidders from $1$ to $i$ and the bid of bidder $i+1$, with $0$
meaning the former is not greater than the latter, $1$ otherwise.
$R^{min}_{1,i}$ denotes the index (starting from 0) of the first
bidder with the minimum bid among the bidders from $1$ to $i$.

%The return values are $R^{min}_{1,n}$ in encrypted form, and the
%corresponding EBV bid $\textbf{e}(v^{n*})$.

It is trivial to use Algorithms \ref{protocol:2mbs} and
\ref{protocol:nmbs} for maximum selection (by inverting the bits of
EBV bids and then calling the minimum selection algorithms). In the
following, we directly use algorithms TwoBidMax(.,.) and
MultiBidMax(.) for maximum selection.

\subsection{Secure Auction Design}

Based on the secure bid representation and operations, we now
present the secure auction design. Our main idea is that the auction
agent first runs the key generation algorithm of Paillier
cryptosystem, and publishes the public key to the auctioneer and the
bidders. Next, all bidders convert their bids to EBV bids using the
public key, and send these EBV bids to the auctioneer. Then, the
auctioneer computes the auction on the EBV bids and gets an
encrypted auction result, with the help of the auction agent.
Finally, the auctioneer gets the auction result by asking the
auction agent to decrypt it, and reports the auction result to the
bidders. As long as the auctioneer and the auction agent do not
collude with each other, they can get nothing about the bids, except
the auction result. PS-TRUST includes three steps as follows.

\subsubsection{Buyer Group Formation}

Buyers submit their location information to the auctioneer, who
generates a conflict graph of buyers based on the information.
Without knowing the bid values of the buyers, the auctioneer forms
buyers into non-conflict buyer groups based on the conflict graph.
Specifically, the auctioneer forms buyer groups by finding
independent sets in the conflict graph repeatedly. To find an
independent set, the auctioneer randomly chooses a node in the
current conflict graph to add to the set, eliminates the node and
its neighbors, and updates the conflict graph. This is repeated
recursively until the conflict graph is empty, then an independent
set is found. We denote by $\mathbb{G} = \{\mathbb{G}_1,
\mathbb{G}_2, ..., \mathbb{G}_H\}$ the set of non-conflict buyer
groups formed.

\subsubsection{Secure Auction Computation}

In this step, all bidders submit their EBV bids to the auctioneer.
Then, the auctioneer and the auction agent cooperate to securely
compute the auction. This step can be divided into further two
steps:

\textbf{(1) Buyer Group Bidding}

For each buyer group $\mathbb{G}_i$ ($1 \le i \le H$), the
auctioneer finds its minimum EBV bid $\textbf{e}(v^{min}_i)$ by
calling $ (R^{min}_{\mathbb{G}_i}, \textbf{e}(v^{min}_i)) =$
MultiBidMin $(E(\mathbb{G}_i))$, and compute its EBV group bid
$\textbf{e}(v^g_i) =$ EBVMul$(\textbf{e}(v^{min}_i), n_i)$, where
$E(\mathbb{G}_i)$ denotes the EBV bid set of group $\mathbb{G}_i$,
and $n_i = |\mathbb{G}_i|$. Note that the auctioneer knows nothing
about the buyers' bids. At the end, the auctioneer holds the EBV
group bid of each buyer group.

\textbf{(2) Winner Determination}

A natural idea for winner determination proceeds as follows. The
auctioneer finds the encrypted seller index (starting from 1)
$E(\alpha)$ with the minimum bid in the seller set $\mathbb{S}$, the
encrypted buyer group index $E(\beta)$ with the maximum group bid in
the buyer group set $\mathbb{G}$, and their corresponding EBV bids
(using Algorithm \ref{protocol:nmbs}), then compares the two EBV
bids to get an encrypted result (using Algorithm
\ref{protocol:2mbs}). $E(\alpha)$, $E(\beta)$, and the comparison
result are sent to the auction agent, who decrypts these encrypted
data, and sends the decrypted information ($\alpha$, $\beta$, and
the comparison result) back to the auctioneer. Then if the trading
condition, namely, the maximum group bid is not less than the
minimum seller bid, is satisfied, the auctioneer removes $\alpha$
from $\mathbb{S}$, $\beta$ from $\mathbb{G}$, and adds $\alpha$ to a
winner-candidate seller set $\mathbb{W}^s$, $\beta$ to a
winner-candidate buyer group set $\mathbb{W}^g$. Otherwise, the
auction is over. This process can be repeated to find the
winner-candidate seller-buyer-group pairs until either the seller
set or the buyer group set is empty, or the trading condition is
unsatisfied. At last, the auctioneer removes the last added seller
$\alpha_c$ from $\mathbb{W}^s$, treating it as the critical seller,
and removes the last added buyer group $\beta_c$ from
$\mathbb{W}^g$, treating it as the critical buyer group. Then, the
auctioneer reports the sellers in $\mathbb{W}^s$ and the buyers
belonging to the buyer groups in $\mathbb{W}^g$ as winners, and the
bid of $\alpha_c$ and the group bid of $\beta_c$ (which are
decrypted by auction agent) as the selling and buying clearing
prices, respectively.

The idea above seems to work well: the auctioneer and the auction
agent cooperate to determine the winners and no exact bids are
leaked to them. However, there is some information about the bids
leaking. Specifically, the ranking orders of the winning sellers'
bids and the winning buyer groups' group bids are leaked to both the
auctioneer and the auction agent. The leaked information is
obviously more than what we can infer from the auction result
including the winner sets and the clearing prices. Thus, in the
sense of cryptography, the above procedure is not really secure.

In order to make this natural procedure of winner determination
secure, something has to be done to hide the bid ranking orders of
winners. Our idea is that, the auctioneer uses the randomized seller
set $\mathbb{S}^{\prime}$ and buyer group set $\mathbb{G}^{\prime}$,
instead of the original ones, so that each time when the auction
agent decrypts the comparison result of a seller-buyer-group pair,
he does not know which the pairs are. The auction agent then
indicates the selected winner-candidate pairs by encrypted bit
vectors, which are sufficient for computing the next
winner-candidate pair while reveals nothing about the selection
orders to the auctioneer. Finally, when the auction is over, the
auction result is decrypted by the auction agent to the auctioneer.
The improved winner determination procedure is depicted in Protocol
\ref{protocol:windeterm}. Some details are explained as follows.

\floatname{algorithm}{Protocol}

\begin{algorithm}[!tb]
\caption{Winner Determination} \label{protocol:windeterm}
\begin{algorithmic}[1]

\REQUIRE ~~\\
    \textbf{Auctioneer (AE)} holds:\\
    EBV bids $\textbf{e}(v^s_i)$ of seller $s_i$, for $1 \le i \le
    M$;\\
    EBV group bids $\textbf{e}(v^g_j)$ of buyer group $\mathbb{G}_j$, for $1 \le j \le
    H$;\\
    Seller set $\mathbb{S} = \{s_1, s_2, ..., s_M\}$;\\
    Buyer group set $\mathbb{G} = \{\mathbb{G}_1, \mathbb{G}_2, ...,
    \mathbb{G}_H\}$.\\

\ENSURE ~~\\
    \textbf{Auctioneer} and \textbf{Auction Agent (AA)} hold:\\
    The selling and buying clearing prices $v^s_c$ and $v^g_c$; \\
    Winning seller set $\mathbb{W}^s$;\\
    Winning buyer group set $\mathbb{W}^g$.\\

\textbf{Step AE1: AE Initialization:}\\

\STATE $\mathbb{W}^s = \emptyset$; $\mathbb{W}^g = \emptyset$;

\STATE $\mathbb{S}^{\prime} = \pi_s(\mathbb{S}) = \{s_{i_1},
s_{i_2}, ..., s_{i_M}\}$;

\STATE $\mathbb{G}^{\prime} = \pi_g(\mathbb{G}) =
\{\mathbb{G}_{j_1}, \mathbb{G}_{j_2}, ..., \mathbb{G}_{j_H}\}$;

\textbf{Step AA2: AA Initialization:}\\

\STATE $\textbf{w}^s = (w^s_1, w^s_2,...,w^s_M )=(0,0,...,0)$;

\STATE $\textbf{w}^g = (w^g_1, w^g_2,...,w^g_H)=(0,0,...,0)$;

\STATE $\alpha_c = -1$; $\beta_c = -1$;

\textbf{Step AE3: Finding a Seller-Buyer-Group Pair:}\\

%\IF{$\mathbb{S}^{\prime} \neq \emptyset$ and $\mathbb{G}^{\prime}
%\neq \emptyset$}

\STATE\label{line:mbmin} $(E(\alpha), \textbf{e}(v^s_{\alpha})) =$
MultiBidMin$(E(\mathbb{S}^{\prime}))$;

\STATE\label{line:mbmax} $(E(\beta), \textbf{e}(v^g_{\beta})) =$
MultiBidMax$(E(\mathbb{G}^{\prime}))$;

\STATE\label{line:tbmax} $(E(R^{max}_{\beta \alpha}),
\textbf{e}(v^{max}_{\beta \alpha})) =$
TwoBidMax$(\textbf{e}(v^g_{\beta}), \textbf{e}(v^s_{\alpha}))$;

\STATE AE sends $E(\alpha)$, $E(\beta)$, and $E(R^{max}_{\beta
\alpha})$ to AA;

\textbf{Step AA4: Determining a Winner-Candidate Pair:}\\

\STATE $\alpha = D(E(\alpha))$; $\beta = D(E(\beta))$;
$R^{max}_{\beta \alpha} = D(E(R^{max}_{\beta \alpha}))$;

\IF{($R^{max}_{\beta \alpha} == 0$)}\label{line:winif1}

\STATE\label{line:plusone1} $w^s_{\alpha + 1} = 1$; $w^g_{\beta + 1}
= 1$;

\STATE $\alpha_c = \alpha$; $\beta_c = \beta$;

\STATE $E(\textbf{w}^s) = (E(w^s_1), E(w^s_2),...,E(w^s_M))$;

\STATE $E(\textbf{w}^g) = (E(w^g_1), E(w^g_2),...,E(w^g_H))$;

\STATE AA sends $E(\textbf{w}^s)$, $E(\textbf{w}^g)$, and
$R^{max}_{\beta \alpha}$ to AE;

\ELSE

\STATE\label{line:plusone2} $w^s_{\alpha_c + 1} = 0$; $w^g_{\beta_c
+ 1} = 0$;

\STATE AA sends $\textbf{w}^s$, $\textbf{w}^g$ and $R^{max}_{\beta
\alpha}$ to AE;

\ENDIF

\textbf{Step AE5: Auction Repeating:}\\

\IF{($R^{max}_{\beta \alpha} == 0$)}\label{line:winif2}

\STATE\label{line:saveebv} $\textbf{e}(v^s_c) =
\textbf{e}(v^s_{\alpha})$; $\textbf{e}(v^g_c) =
\textbf{e}(v^g_{\beta})$;

\STATE\label{line:updates} Computes $\textbf{e}(v^s_{i_k}) =
(E(\sigma^{s*}_{i_k,1}), E(\sigma^{s*}_{i_k,2}), ...,
E(\sigma^{s*}_{i_k,K}))$, where $\sigma^{s*}_{i_k,p} =
\sigma^s_{i_k,p} + w^s_k \cdot(1 - \sigma^s_{i_k,p})$, $1 \le p \le
K$, for all $1 \le k \le M$;

\STATE\label{line:updatebg} Computes $\textbf{e}(v^g_{j_k}) =
(E(\sigma^{g*}_{j_k,1}), E(\sigma^{g*}_{j_k,2}), ...,
E(\sigma^{g*}_{j_k,K}))$, where $\sigma^{g*}_{j_k,p} =
\sigma^g_{j_k,p} - w^g_k \cdot \sigma^g_{j_k,p}$, $1 \le p \le K$,
for all $1 \le k \le H$;

\STATE Goto \textbf{Step AE3};

\ENDIF

\textbf{Step AE6: Auction Opening:}\\

\STATE AE gets $v^s_c$ and $v^g_c$ by asking AA to decrypt
$\textbf{e}(v^s_c)$ and $\textbf{e}(v^g_c)$;

\STATE $\mathbb{W}^s = \{s_{i_k} | w^s_k = 1, \forall 1 \le k \le
M\}$;

\STATE $\mathbb{W}^g = \{\mathbb{G}_{i_k} | w^g_k = 1, \forall 1 \le
k \le H\}$;

\end{algorithmic}
\end{algorithm}

In Step AE1, the auctioneer applies random permutations $\pi_s$ and
$\pi_g$ to seller set $\mathbb{S}$ and buyer group set $\mathbb{G}$,
respectively, getting the randomized sets $\mathbb{S}^{\prime}$ and
$\mathbb{G}^{\prime}$. Note that only the auctioneer knows the
permutations.

In Step AA2, two bit vectors $\textbf{w}^s$ and $\textbf{w}^g$ are
defined to indicate the winner locations in the randomized sets
$\mathbb{S}^{\prime}$ and $\mathbb{G}^{\prime}$, respectively.
$w^s_k = 1$ if seller $s_{i_k}$ is a candidate winner, $w^s_k = 0$
otherwise, and $w^g_k = 1$ if buyer group $\mathbb{G}_{j_k}$ is a
candidate winner, $w^g_k = 0$ otherwise. $\alpha_c$ and $\beta_c$
index the critical seller and buyer group, respectively.

In Step AE3, similarly to the natural idea, the encrypted seller
index $E(\alpha)$ with the minimum bid and the encrypted buyer group
index $E(\beta)$ with the maximum bid, together with their EBV bids
are computed using Algorithm \ref{protocol:nmbs}. The resulted two
EBV bids are then compared using Algorithm \ref{protocol:2mbs}.
These computation results remain in the encrypted form, unknown to
the auctioneer. Note that, different from the natural idea, the
randomized sets $\mathbb{S}^{\prime}$ and $\mathbb{G}^{\prime}$ are
used instead.

In Step AA4, the auction agent decrypts the computation results in
Step AE3, knowing the locations of the candidate-winner pair in the
randomized sets  $\mathbb{S}^{\prime}$ and $\mathbb{G}^{\prime}$.
However, he does not know the random permutations, so he cannot know
the true candidate winners. Line \ref{line:winif1} tests if the
buyer group's bid is not less than the seller's bid. If so, the
auction agent sets the corresponding bits of $\textbf{w}^s$ and
$\textbf{w}^g$ to 1, saves indexes of the last candidate-winner
pair, and sends $E(\textbf{w}^s)$, $E(\textbf{w}^g)$ and
$R^{max}_{\beta \alpha}$ to the auctioneer. Otherwise, the auction
is over, and auction agent removes the last candidate-winner pair
(i.e. the critical seller $\alpha_c$ and buyer group $\beta_c$) from
candidate winner sets by setting the corresponding bits of
$\textbf{w}^s$ and $\textbf{w}^g$ to 0. The auction agent then sends
the plain values including $\textbf{w}^s$, $\textbf{w}^g$ and
$R^{max}_{\alpha \beta}$ to the auctioneer.

In Step AE5, Line \ref{line:winif2} tests if seller $\alpha$ and
buyer group $\beta$ can be included to the winner-candidate sets. If
so, the auctioneer first saves the EBV bids of the last
winner-candidate pair in Line \ref{line:saveebv}, and then updates
the EBV bids of all sellers and all buyer groups in Lines
\ref{line:updates} and \ref{line:updatebg}, respectively. This
updating results that the bid of seller $s_{i_k}$ is set to $(2^K -
1)$ if $w^s_k == 1$, while remains unchanged otherwise, and the bid
of buyer group $\mathbb{G}_{j_k}$ is set to $0$ if $w^g_k == 1$,
while remains unchanged otherwise. That is, all selected
winner-candidate sellers are mapped to a maximum value $(2^K - 1)$,
and all selected winner-candidate buyer groups are mapped to a
minimum bid value $0$. As long as the normal bid satisfies $0 < v <
2^K-1$, the selected winner candidates will not be selected in Step
AE3, and the updating is equivalent to removing the winner
candidates from the seller set and buyer group set. After doing this
updating, the execution goes to Step AE3. If the test of Line
\ref{line:winif2} fails, the auction repeating is over and the
execution goes to Step AE6.

In step AE6, the auctioneer gets $v^s_c$ and $v^g_c$ by asking the
auction agent to decrypt $\textbf{e}(v^s_c)$ and
$\textbf{e}(v^g_c)$, and computes the winner sets $\mathbb{W}^s$ and
$\mathbb{W}^g$ from $\textbf{w}^s$ and $\textbf{w}^g$ using the
randomization permutations in Step AE1.

Note that in Line \ref{line:plusone1} and \ref{line:plusone2} in
Protocol \ref{protocol:windeterm}, the need of ``adding one'' is
caused by different ways of indexing, i.e., $\alpha$ and $\beta$
returned by Algorithms \ref{protocol:2mbs} or \ref{protocol:nmbs}
are starting from 0, while the indexes of sellers and buyer groups
are from 1.

\subsubsection{Pricing}

Each spectrum channel is sold from the winning sellers at the
selling clearing price $v^s_c$, and bought by the winning buyer
groups at the buying clearing price $v^g_c$. Each winner buyer in
winning buyer group $\mathbb{G}_k$ pays the equal share of the
buying clearing price, that is $v^g_c/n_k$, where $n_k =
|\mathbb{G}_k|$.

From the description above, we can see that PS-TRUST exactly follows
the auction procedure of TRUST. Therefore, PS-TRUST maintains the
properties of economic-robustness and spectrum reuse of TRUST, in
the presence of semi-honest adversaries.

\subsection{Security Analysis}

In the sense of cryptography, the standard definition of security
against semi-honest adversaries can be described as follows
\cite{04goldreicho}.

\begin{mydef}[Security against Semi-honest Adversaries]
Let $f(x,y)$ be a functionality with two inputs $x$ and $y$, and two
outputs $f^A(x,y)$ and $f^B(x,y)$. Suppose that protocol $\Pi$
computes functionality $f(x,y)$ between two parties Alice and Bob.
Let $V^{\Pi}_A(x,y)$ (resp. $V^{\Pi}_B(x,y)$) represent Alice's
(resp. Bob's) view during an execution of $\Pi$ on $(x,y)$. In other
words, if $(x, \textbf{r}^{\Pi}_A)$ (resp. $(y,\textbf{r}^{\Pi}_B)$)
denotes Alice's (resp. Bob's) input and randomness, then
\[
\begin{array}{l}
V^{\Pi}_A(x,y) = (x, \textbf{r}^{\Pi}_A, m_1, m_2,...,m_t),\text{ and} \\
V^{\Pi}_B(x,y) = (y, \textbf{r}^{\Pi}_B, m_1, m_2,...,m_t)
\end{array}
\]
where $\{m_i\}$ denote the messages passed between the parties. Let
$O^{\Pi}_A$ (resp. $O^{\Pi}_B$) denote Alice's (resp. Bob's) output
after an execution of $\Pi$ on $(x,y)$, and $O^{\Pi}(x,y) =
(O^{\Pi}_A(x,y), O^{\Pi}_B(x,y))$. Then we say that protocol $\Pi$
is secure (or protects privacy) against semi-honest adversaries if
there exist probabilistic polynomial time (PPT) simulators $S_1$ and
$S_2$ such that
\begin{equation}\label{equ:semisecure1}
\{(S_1(x, f_A(x,y)), f(x,y))\} \overset{c}{\equiv}
\{(V^{\Pi}_A(x,y), O^{\Pi}(x,y))\}
\end{equation}
\begin{equation}\label{equ:semisecure2}
\{(S_2(x, f_B(x,y)), f(x,y))\} \overset{c}{\equiv}
\{(V^{\Pi}_B(x,y), O^{\Pi}(x,y))\}
\end{equation}
where $\overset{c}{\equiv}$ denotes computational
indistinguishability.
\end{mydef}

With the above security definition, we now prove the basic lemma
that will allow us to argue that our auction solution is secure
against semi-honest adversaries. Lemma \ref{lem:lemma1} is similar
to Lemma 1 in \cite{07Bunnp}, with slight difference and some
extension.

\begin{mylemma}\label{lem:lemma1}
Suppose that Alice has run the key generation algorithm for
semantically secure homomorphic public-key encryption scheme, and
has given her public key to Bob. Suppose also that Alice and Bob run
Protocol $X$, for which all messages passed from Alice to Bob are
encrypted using this scheme, or only carry information that can be
completely inferred from the output of Bob, and all messages passed
from Bob to Alice are uniformly distributed in their value ranges
and independent of Bob's inputs, or only carry information that can
be completely inferred from the output of Alice. Then Protocol $X$
is secure against semi-honest adversaries.
\end{mylemma}

\begin{proofname}
We prove the security of Protocol X in two separate cases, depending
on which party the adversary has corrupted. To prove security, we
show that for all PPT adversaries, the adversary's view based on
Alice and Bob's interaction is indistinguishable to the adversary's
view when the corrupted party interacts with a simulator instead. In
other words, we show that there exist simulators $S_1$ and $S_2$
that satisfy conditions (\ref{equ:semisecure1}) and
(\ref{equ:semisecure2}).

Case 1: Bob is corrupted. We simulate Alice's messages sent to Bob.
For each encrypted message that Alice is supposed to send to Bob, we
let the simulator $S_2$ pick a random element from $\mathbb{Z}_N$,
and send an encryption of this. Any adversary who can distinguish
between interaction with Alice versus interaction with $S_2$ can be
used to break the security assumptions of the used encryption
scheme. Thus, no such PPT adversary exists. For each (plain) message
that only carries information that can be completely inferred from
the output of Bob, the simulator $S_2$ can of course simulate it
using Bob's output of the functionality ($f_B(x,y)$ in equation
(\ref{equ:semisecure2})).Thus, condition (\ref{equ:semisecure2})
holds.

Case 2: Alice is corrupted. We simulate Bob's messages sent to
Alice. For each message that is uniformly distributed in its value
range and independent of Bob's inputs, simulator $S_1$ picks a
random element from its range and sends to Alice. Again, equation
(\ref{equ:semisecure1}) holds due to the fact that Alice cannot
distinguish the simulator's random element from the correct element
that has been randomized by Bob over its value range. For each
message that only carries information that can be completely
inferred from the output of Alice, the simulator $S_1$ can simulate
it using Alice's output of the functionality ($f_A(x,y)$ in equation
(\ref{equ:semisecure1})). Thus, condition (\ref{equ:semisecure1})
holds.

Thus, we can conclude that Protocol $X$ is secure against
semi-honest adversaries. $\Box$
\end{proofname}

\begin{mytheorem}\label{the:product}
Protocol \ref{protocol:product} is secure against semi-honest
adversaries.
\end{mytheorem}

\begin{proofname}
It is obvious that all messages passed from AR to KH are uniformly
distributed in the ciphertext space $\mathbb{Z}_{N^2}$ (or the
values obtained by decrypting the messages are uniformly distributed
in the plaintext space $\mathbb{Z}_{N}$), and the messages passed
from KH to AR are encrypted. According to Lemma \ref{lem:lemma1},
Protocol \ref{protocol:product} is secure against semi-honest
adversaries. $\Box$
\end{proofname}

\begin{mytheorem}\label{the:alg}
Suppose that the auction agent has run the key generation algorithm
for semantically secure homomorphic public-key encryption scheme,
and has given its public key to the auctioneer. Further suppose that
the auctioneer runs Algorithm $X$ (where $X$ is one of
\ref{protocol:add}, \ref{protocol:mul}, \ref{protocol:2mbs},
\ref{protocol:nmbs}), and holds the computation result. Then the
resulting protocol is secure against semi-honest adversaries.
\end{mytheorem}

\begin{proofname}
The resulting protocol has no messages exchanged, except
sequentially calling Protocol \ref{protocol:product} which is secure
against semi-honest adversaries, so due to Lemma \ref{lem:lemma1}
and sequential composition theory \cite{10Hazayc}, it is secure
against semi-honest adversaries.$\Box$
\end{proofname}

\begin{mytheorem}\label{the:windeterm}
Protocol \ref{protocol:windeterm} is secure against semi-honest
adversaries.
\end{mytheorem}

\begin{proofname}
We show that all the messages exchanged between the parties satisfy
the conditions of Lemma \ref{lem:lemma1}. Then, applying Lemma
\ref{lem:lemma1} and the sequential composition theory
\cite{10Hazayc}, Protocol \ref{protocol:windeterm} is secure against
semi-honest adversaries.

Specifically, suppose that there are $Q$ winner-candidate pairs
(including the critical seller and buyer group), we can list all the
messages exchanged between the parties as follows.

Messages sent from AE to AA include:
\[
\{E(\alpha_i)\}^{Q+1}_1, \{E(\beta_i)\}^{Q+1}_1,
\{E(R^{max}_{\beta_i \alpha_i})\}^{Q+1}_1, \textbf{e}(v^s_c),
\textbf{e}(v^g_c)
\]

Message sent from AA to AE include:
\[
\{E(\textbf{w}^s_i)\}^Q_1, \{E(\textbf{w}^g_i)\}^Q_1,
\textbf{w}^s_{Q+1}, \textbf{w}^g_{Q+1}, \{R^{max}_{\beta_i
\alpha_i}\}^{Q+1}_1, v^s_c, v^g_c
\]

Now we show that all these messages satisfy the conditions of Lemma
\ref{lem:lemma1}. First, among the messages sent from AE to AA,
$\alpha_i$ and $\beta_i$ (obtained by decrypting $E(\alpha_i)$ and
$E(\beta_i)$) are uniformly distributed over their value ranges
(i.e. $[1..M]$ and $[1..H]$) due to the random permutations unknown
to AA, and messages $R^{max}_{\beta_i \alpha_i}$, $v^s_c$ and
$v^g_c$ can be completely inferred from the output of AA, which is
also the auction result including selling and buying clearing prices
$v^s_c$, $v^g_c$, and the winner sets $\mathbb{W}^s$ and
$\mathbb{G}^g$. Second, among the messages sent from AA to AE,
messages $E(\textbf{w}^s_i)$ and $E(\textbf{w}^g_i)$ are encrypted,
and messages $\textbf{w}^s_{Q+1}$, $\textbf{w}^g_{Q+1}$,
$R^{max}_{\beta_i \alpha_i}$, $v^s_c$ and $v^g_c$ can be completely
determined by the output of the auctioneer, which is also the
auction result. As a result, all the messages in Protocol
\ref{protocol:windeterm} satisfy the conditions of Lemma
\ref{lem:lemma1}.

Furthermore, according to Theorem \ref{the:alg}, subprotocols
resulted from running Algorithms \ref{protocol:2mbs} and
\ref{protocol:nmbs} (i.e. calling MultiBidMin$(.)$, MultiBidMax$(.)$
and TwoBidMax$(., .)$) are secure against semi-honest adversaries.
Then, applying Lemma \ref{lem:lemma1} and sequential composition
theory, we can conclude that Protocol \ref{protocol:windeterm} is
secure against semi-honest adversaries. $\Box$

\end{proofname}

Now, we can conclude PS-TRUST is secure against semi-honest
adversaries.

\begin{mytheorem}
PS-TRUST is a two-party protocol secure against semi-honest
adversaries, between the auctioneer and the auction agent.
Additionally, anyone (i.e. the auctioneer, auction agent, and each
bidder) cannot know anything about the bids beyond the auction
result through the auction.
\end{mytheorem}

The proof is obvious based on the previous theorems, and we only
sketch it here. Note that in the auction, we implicitly assume that
the bidders' bids are the only privacy needed to protect. So, steps
of Buyer Group Formation and Pricing of the auctions are unrelated
to the security. That is, we only need to prove that the step of
Secure Auction Computation is secure. By Theorem
\ref{the:windeterm}, the winner determination procedure is secure,
and we can similarly prove the security of buyer group bidding
procedure. Thus, PS-TRUST is secure against semi-honest adversaries.
What is more, because bidders' bids are encrypted in EBV form, and
are input to the auctioneer, according to the definition of
security, neither the auctioneer nor the auction agent knows
anything about the bids, and no bidder knows anything about other
bidders' bids, except the auction result.

\section{Performance Analysis and Evaluation}\label{sec:experiment}

As PS-TRUST exactly follows the procedure of TRUST, the auction
efficiency is the same as that of TRUST. So, we only focus on the
analysis and evaluation of computation and communication overhead
caused by the security measures.

\subsection{Performance Analysis}

The analysis of computation and communication complexities for
Protocols/Algorithms from \ref{protocol:product} to
\ref{protocol:nmbs} is straightforward and the results are listed in
Tab. \ref{tab:comput}. We thus can find the computation complexity
of Protocol \ref{protocol:windeterm} (which is also the computation
complexity PS-TRUST) is $O((M + N) \cdot K \cdot W)$ operations
(e.g. addition or multiplication) of big integers, where $W$
represents the number of seller-buyer-group winner pairs. Similarly,
we can find the communication complexity of PS-TRUST is $O((M + N)
\cdot K \cdot W)$ times of bit length of big integers. Note that,
practical running time and message volume will be impacted by the
bit length used in the homomorphic encryption scheme.

\begin{table}[!t]
\renewcommand{\arraystretch}{1.3}

\caption{Computation and Communication Complexities}
\label{tab:comput} \centering
% Some packages, such as MDW tools, offer better commands for making tables
% than the plain LaTeX2e tabular which is used here.
\begin{tabular}{|c||c||c||c||c||c|}
\hline
Protocol/Algorithm & \ref{protocol:product} & \ref{protocol:add} & \ref{protocol:mul} & \ref{protocol:2mbs} & \ref{protocol:nmbs} \\
\hline
Complexity & $O(1)$ & $O(K)$ & $O(K^2)$ & $O(K)$ & $O(nK)$\\
\hline
\end{tabular}
\end{table}

\subsection{Performance Evaluation}

We implement PS-TRUST using Java in Windows XP with Intel's Core 2
Duo CPU 2.93GHz. We let the buyers be randomly distributed in an
area of $100\text{m} \times 100\text{m}$, let the protection
distance be $50$m, and let default experimental setting be as
follows: the bit length of homomorphic encryption scheme is 512,
i.e., $N$'s bit length is 512; the bit length $K$ of EBV is 8; the
numbers $(M, N)$ of sellers and buyers are $(10, 30)$. All
experimental results are averaged on 10 random repetitions.

\begin{figure}[ht]
\centering
\includegraphics[width=0.5\textwidth]{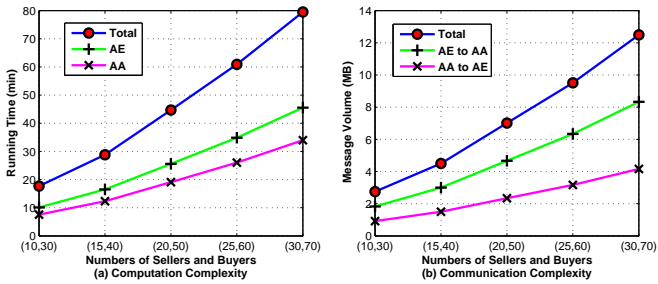}
\caption{Overhead Evaluation as the Numbers of Sellers and Buyers
Vary} \label{fig:experiment}
\end{figure}

Fig. \ref{fig:experiment} shows the curves of running times and
message volumes of PS-TRUST as $(M, N)$ vary from $(10, 30)$ to
$(30, 70)$. Both performance measures grow slightly faster than
linear growth according to $(M + N)$. This is because according to
the theoretical results, these measures also depend on $W$, which
increases  as well with $(M + N)$ on average.

\begin{figure}[ht]
\centering
\includegraphics[width=0.5\textwidth]{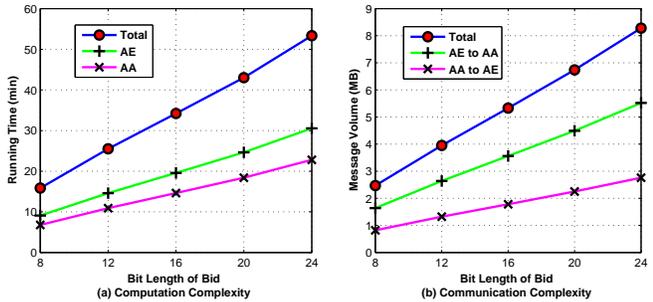}
\caption{Overhead Evaluation as the Bit Length of Bids Varies}
\label{fig:experiment2}
\end{figure}

Fig. \ref{fig:experiment2} show the curves of running times and
message volumes of PS-TRUST as $K$ vary from 8 to 24. We can see
that all the curves are roughly linear to $K$. This is consistent
with the theoretical results fairly well.

From the analytical and experimental results above, we can see that
both running times and message volumes are feasible for practical
applications. Furthermore, the running time of the auctioneer (AE)
is about a third more than that of the auction agent (AA), and the
message volume of AE is about twice of that of the AA. Finally, the
running times can be reduced by parallel computing if needed.

\section{Conclusion}\label{sec:conclusion}
In this paper, we have proposed PS-TRUST, the first provably secure
solution to truthful double spectrum auctions. Previous studies on
secure spectrum auctions did not provide adequate security, as they
revealed information about the bids beyond the auction result.
Different from those studies, we have achieved security in the sense
of cryptography in this work. Specifically, PS-TRUST reveals nothing
about the bids to any participant, except the auction result
including clearing prices and winner sets. We have also proved
formally the security of PS-TRUST in the presence of semi-honest
adversaries. Finally, we have implemented PS-TRUST in Java, and have
theoretically and experimentally shown that the computation and
communication overhead of PS-TRUST is modest, and its practical
applications are feasible.


%\section*{Acknowledgment}



%\bibliographystyle{IEEEtran}

\end{document}